\documentclass[a4paper,twocolumn,11pt,unpublished]{quantumarticle}

\pdfoutput=1
\usepackage[utf8]{inputenc}
\usepackage[english]{babel}
\usepackage[T1]{fontenc}
\usepackage{amsmath}
\usepackage{tikz}
\usepackage{fancybox}

\usepackage{hyperref}
\usepackage[numbers]{natbib}
\begin{document}

\title{Vertical ion transport in a surface Paul trap: escalator and elevator approaches}

\author{Alexey Russkikh}
\affiliation{P.N. Lebedev Physical Institute, Russian Academy
of Sciences, Moscow, Russia}
\author{Nikita Zhadnov}
\email{nik.zhadnov@gmail.com}
\orcid{0000-0002-9747-3028}
\affiliation{P.N. Lebedev Physical Institute, Russian Academy
of Sciences, Moscow, Russia}
\affiliation{Russian Quantum Center, Moscow, Russia}
\maketitle

\begin{abstract}
Surface ion traps confining and manipulating tens of ion 
qubits have become the leading platform for quantum processors 
with high quantum volume. These devices employ the Quantum 
Charge-Coupled Device (QCCD) architecture, wherein multiple 
trapping zones are linked by an on-chip transport network that 
shuttles ion chains, enabling full connectivity through physical 
ion transport in a plane parallel to the chip surface. The 
ability to move ions perpendicular to this plane can offer 
additional advantages, including tuning the laser--ion 
interaction strength, systematic studies of surface-induced 
heating mechanisms, and precise alignment with a mode of an 
external optical cavity. We introduce an ``escalator''---a 
geometrically optimized transition between trapping zones of 
different confinement heights---and present a comparative 
analysis of two ``elevator'' configurations that reposition the 
RF null dynamically via additional electrode voltages. Both approaches enable nearly a twofold change in the ion confinement height above the chip surface.
\end{abstract}

\section{Introduction}
Recent success in scaling ion quantum computers \cite{kolachevsky2025quantum} has been attributed to early implementations of the QCCD architecture \cite{kielpinski2002architecture}, namely, its modular trap layout, gate operations enabled by ion shuttling and functionally differentiated trap zones \cite{moses2023race,ransford2025helios}. These techniques allowed the demonstration of the most complicated quantum computations to date \cite{DeCross2025computational}. Surface Paul traps \cite{chiaverini2005surface} are particularly well suited to the QCCD architecture, as they readily support complicated networks of connected trapping regions and permit controlled transport of ion chains across a planar substrate. 


To date, many studies have focused on modeling and experimentally implementing the transport of ions in a plane parallel to the surface of the chip, including shuttling the ions along the trap axis \cite{bowler2012coherent,walther2012controlling}, guiding them through junctions \cite{kaushal2020shuttling,sterk2024multi}, and swapping their positions \cite{kaufmann2017fast}. 
In this work, we turn our attention to vertical (out‐of‐plane) transport.
Studies of out‐of‐plane ion transport in surface Paul traps were performed for point traps \cite{kim2010surface}, linear traps \cite{van2016integrated,boldin2018measuring}, and vertical-linear traps \cite{an2018surface}.
Introducing this third spatial degree of freedom offers several advantages:
\begin{enumerate}
  \item \textbf{Control of coupling to addressing lasers and microwave fields:} Vertical transport allows ions to be selectively moved into or out of the focus of global laser beams, enabling zone-specific operations (e.g., memory vs. interaction). Similarly, bringing an ion closer to the chip surface significantly enhances its interaction with integrated microwave antennas, facilitating high-fidelity quantum gates \cite{hughes2025trapped}. 
  \item \textbf{Alignment with an external optical cavity mode:} Precise positioning is essential for maximizing the coupling between an ion and the mode of an external optical cavity \cite{kassa2018precise,takahashi2020strong}, a critical requirement for photonic interconnects in quantum networks and multi-core processors.
  Furthermore, such coupling enables quantum operations between distinct ion chains (located on the same chip or on different ones) via a common cavity mode \cite{ramette2022any}, which fundamentally requires aligning several ions along the cavity axis.
  Since these cavity modes are typically oriented parallel to the chip surface, precise vertical control of the ion's position is indispensable.
  The ion must be placed at the central antinode of the optical standing wave with sub-wavelength accuracy to maximize the coherent coupling rate, which depends on the local electric field strength.
  \item \textbf{Surface noise characterization and mitigation:} The ability to vary the ion-surface distance in situ provides a direct method to probe electric-field noise and study its sources \cite{brownnutt2015ion}.  Increasing the confinement height (e.g. of a memory zone) can be used to mitigate motional decoherence.
\end{enumerate}

The confinement height in a surface ion trap is determined primarily by the widths of the RF electrodes (usually designated $b$ and $c$) and the ground electrode ($a$) \cite{house2008analytic}.
Although DC voltages can displace the ion, this typically introduces excess micromotion and associated heating \cite{leibfried2003quantum}.
Therefore, moving an ion vertically requires shifting the minimum of the radio-frequency pseudopotential (RF null) itself.

Several methods for vertical ion transport have been demonstrated theoretically and experimentally.
One primary approach involves the dynamic control of RF potentials to shift the RF null. This can be achieved by applying a controllable RF voltage to the central electrode, allowing variation of the ion-surface distance, $h$, over a wide range of $50$–$150\,\mu$m \cite{boldin2018measuring}. 
A similar principle of controlling RF potentials on segmented central electrodes or RF rails has also been shown to allow vertical positioning, facilitating ion movement on the scale of tens of micrometers \cite{van2016integrated,channa2019surface}.
An alternative method employs a centrally symmetrical RF electrode configuration to establish a principal trapping axis perpendicular to the chip surface \cite{an2018surface}. In this design, vertical control was achieved across a range of $50$–$300\,\mu$m using DC electrodes.

\begin{figure}[t]
  \centering
  \includegraphics[height=11 cm]{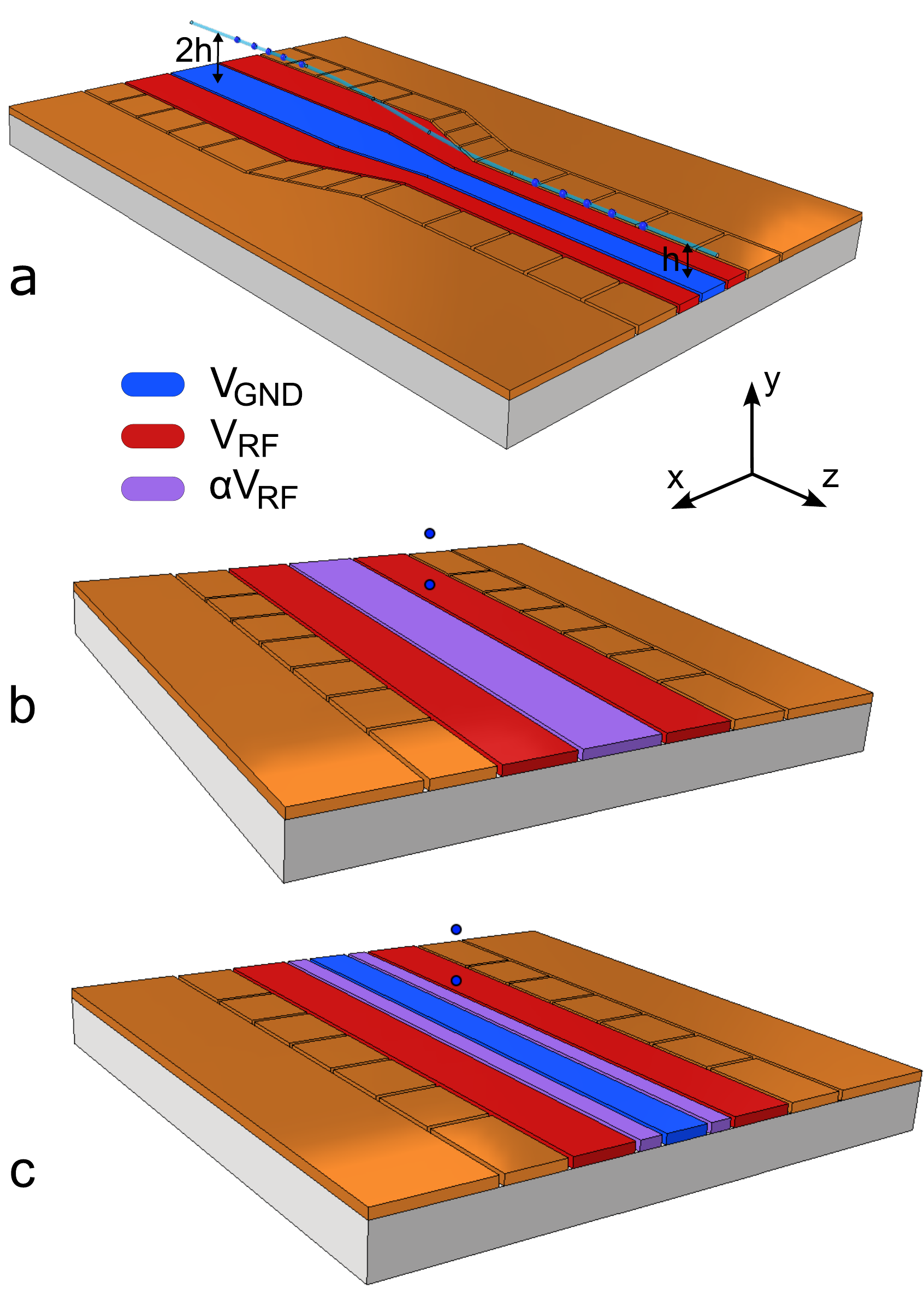}
  \caption{Surface ion trap configurations for vertical ion positioning. (a)~Escalator: a transition region connects two trapping zones with a twofold height difference; the blue line shows the ion trajectory during transport. (b)~Elevator with a controllable RF voltage $\alpha V_\text{rf}$ applied to the central electrode. (c)~Elevator with a controllable RF voltage $\alpha V_\text{rf}$ applied to segments of the central electrode. Two ion positions corresponding to different control voltages are shown in (b) and (c).}
  \label{fig:fig1}
\end{figure}

This work explores a new method  for vertical ion transport in surface traps based on connecting two trapping regions with different inherent confinement heights. This approach does not require additional voltage control and geometrically divides the ion chip into several zones with different levels of confinement. We also consider a known technique of applying additional RF voltages to the central electrode or its segments. We designate the first method an ``escalator'' and the second an ``elevator''. Both concepts are illustrated in Fig.~\ref{fig:fig1}. Our primary goal is to explore the physics and tradeoffs of vertical ion transport rather than to prescribe a specific trap layout, which will necessarily depend on the 
intended application.

\section{Ion chip escalator}

The trap design, shown in Fig.~\ref{fig:fig1}~(a), does not correspond to the five-wire symmetry of a conventional linear surface trap, introducing potential barriers and distortions in the common Paul-trap pseudopotential.
A similar problem occurs for other surface linear Paul trap features, such as loading holes and various types of junctions.
The problem can usually be resolved by carefully shaping the electrode.
The standard strategy for planar junctions is to replace the straight boundary between the central ground electrode and the RF rails with a segmented optimized contour, determined by numerical pseudopotential simulations, to ensure smooth low-heating ion transport \cite{Liu2014,wright2013reliable,zhang2022optimization}.  
In the following sections, we extend this framework to design and optimize an ``escalator'' junction that adiabatically transports ions across a twofold height difference with minimal excitation.

\subsection{Initial design of ion escalator}
In order to create a connected pair of traps with a height difference, we first selected the dimensions of the traps' electrodes and the length of the connection.
In our lab, we work with the $^{171}$Yb$^+$ ion. In all cases, unless otherwise specified, the calculation will be given for this ion with the trap RF voltage parameters $V_\text{rf}$ = 100 V and $\omega_{rf}=2\pi \times 20\,$MHz.  

A trap designed with a smaller height was chosen to meet basic requirements for quantum computing with trapped ions \cite{gerasin2024optimized}. The electrode widths are $80\,\mu$m for the central GND electrode and $b_1 = 65\,\mu$m for RF-electrodes. This design gives radial secular frequency $\omega_{sec} \approx 2.4$ MHz, ion-to-surface distance $h_1 = 71 \ \mu m$ and trap depth $\approx 70$ meV. These parameters are quite common for surface ion traps.

\begin{figure}[!ht]
    \centering
    \includegraphics[width=0.47\textwidth]{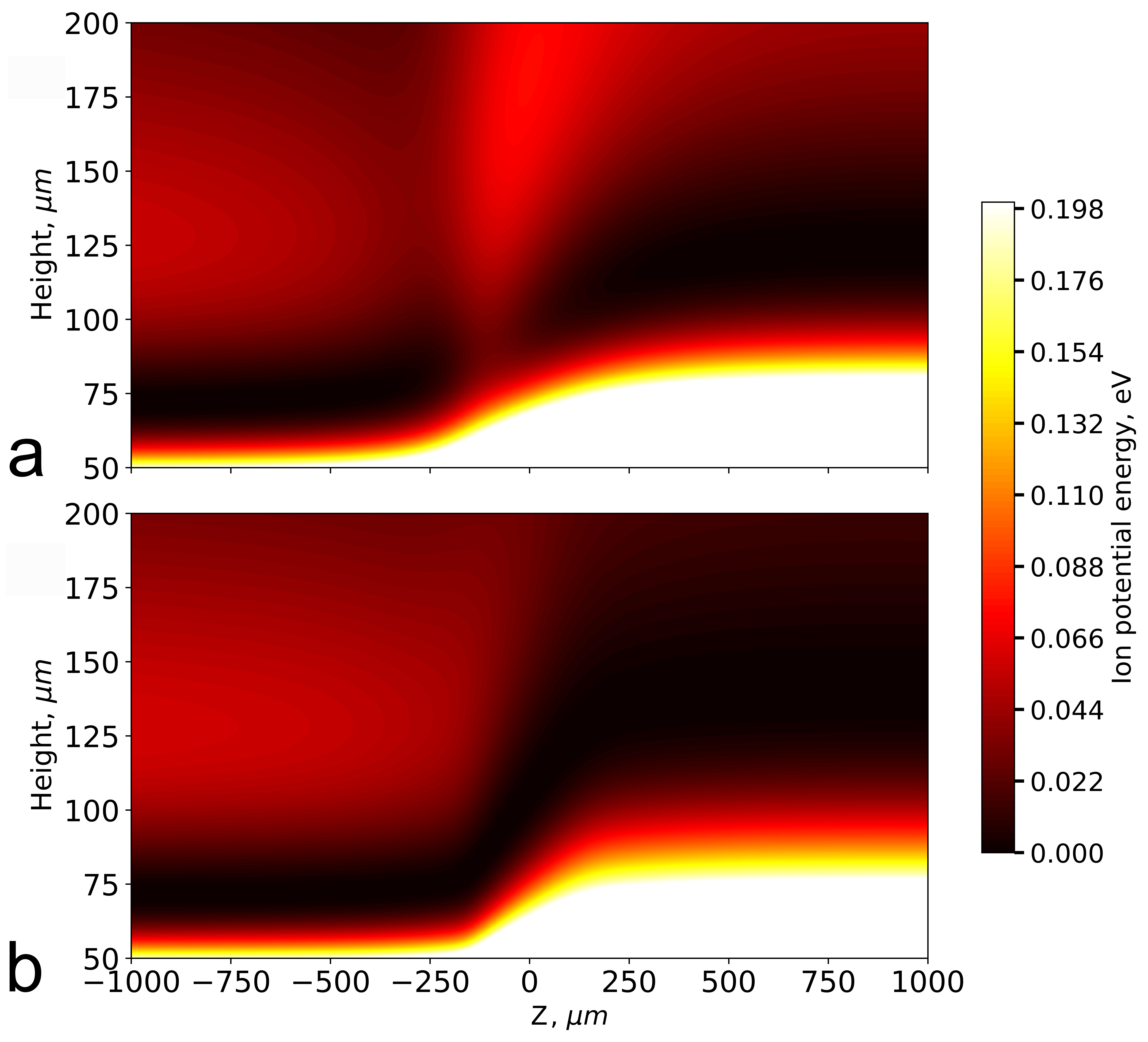}            
    \caption{Pseudopotential distribution along the transport axis. (a)~Unoptimized connection between two trapping zones of different heights. (b)~Optimized transition after all optimization procedures, with a tenfold reduction in the pseudopotential barrier.}
    \label{fig:equi}
\end{figure}

The breaking of translational symmetry near the connection zone between traps creates pseudopotential barriers along the ion shuttling path, as illustrated in Fig.~\ref{fig:equi}~(a).
The second design step optimized the electrode geometry of the second trap and the connection width (see Appendix A). 
The selected connector width is $D = 300\ \mu\text{m}$. For the optimized second trap, the resulting pseudopotential parameters are a confinement height of $141\,\mu$m, radial secular frequencies of $\omega_s \approx0.5\ $MHz, and a trap depth of $\approx 22\,\text{meV}$. 

\subsection{Optimization of connector}
\begin{figure}[!ht]
    \centering
    \includegraphics[width=0.5\textwidth]{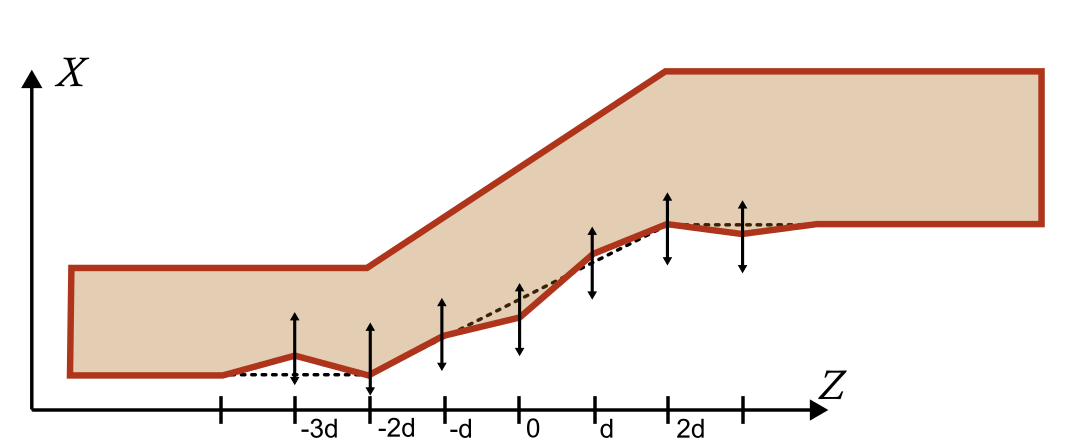}            
    \caption{Variable points of an RF electrode along the transition region.
During optimization, the x-coordinate of each point is varied while keeping z fixed. The points of the lower RF electrode are constructed symmetrically. The DC electrode is adjusted to maintain a constant gap between adjacent electrodes.}
    \label{fig:varik}
\end{figure}

Transporting an ion through pseudopotential barriers excites its motional state, increasing its vibrational quantum number $n$ and reducing the fidelity of transport.
To prevent this, we construct a multi-objective function \cite{Liu2014} that depends on the shape of the electrodes and minimize it.
The optimization goals include not only decreasing the pseudopotential values, but also minimizing its gradient along the ion's path.
This is critical because the resulting heating rate is proportional to $\left(\frac{\partial \psi}{\partial z}\right)^2$ and the RF voltage noise spectral density at secular frequencies $S_{V_{\mathrm{N}}}\left(\Omega_{\mathrm{rf}}-\omega_z\right)$ 
\cite{blakestad2009high,brownnutt2015ion}. By minimizing the potential gradient, we reduce the sensitivity of the ion to this noise source.
To achieve this, we define four specific objective functions that quantify different aspects of the pseudopotential landscape.
\begin{equation}
\begin{aligned}
    &F_1 = \int_{-l}^{l}\psi(y|_{\psi_{min}})dz & \\
    &F_2 = \max\left(\psi(y|_{\psi_{min}})\right) & \\
    &F_3 = \int_{-l}^{l}\left|\frac{\partial\psi(y|_{\psi_{min}})}{\partial z}\right|dz & \\
    &F_4 = \max\left(\left|\frac{\partial\psi(y|_{\psi_{min}})}{\partial z}\right|\right), & \\
\end{aligned}
    \end{equation}
where $\psi$ is pseudopotential, $l = 1000\ \mu$m defines the size of calculation domain and  $y|_{\psi_{\min}}$ denotes the vertical coordinate of the pseudopotential minimum in the $x$--$y$ plane at a given $z$.
$F_1$ and $F_3$ capture the integrated (average) pseudopotential and gradient, while $F_2$ and $F_4$ capture their peak values.
Using these four functions, a multi-objective function is constructed:
\begin{equation*}
    F_0 = \sum_{i=1}^4 \sigma_i\frac{F_i}{F_{i}^\text{norm}},
\end{equation*}
where $\sigma_i$ are the weighting coefficients, $F_{i}^{\text{norm}}$ are the normalization factors. Typically, the normalization constants are chosen as values obtained after minimizing each corresponding function $F_i$. The weights were manually selected based on the relative importance of each objective function to improve overall performance.
The optimized parameters consist of x-coordinates for a set of points along the RF and DC electrode boundaries within the range from $-\text{D}$ to $\text{D}$ (where D is the transition zone width), as shown in Fig.~\ref{fig:varik}. 

    


The simulation parameters $V_{peak}$, $\omega_{rf}$, the ion mass and charge match those described above. The electrode boundary was discretized into 34 control points (spaced at $\approx9 \ \mu$m  intervals), easily compatible with photolithography capabilities. 

\begin{figure}[!h]
    \centering
    \includegraphics[width=0.47\textwidth]{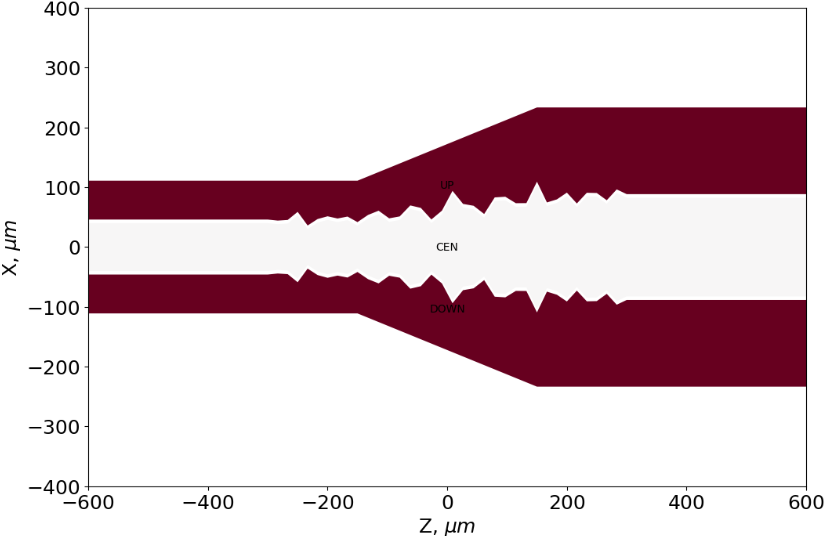}            
    \caption{Optimized electrode geometry of the escalator transition region.}
    \label{fig:dizain}
\end{figure}

\begin{figure}[h!]
    \includegraphics[width=0.47\textwidth]{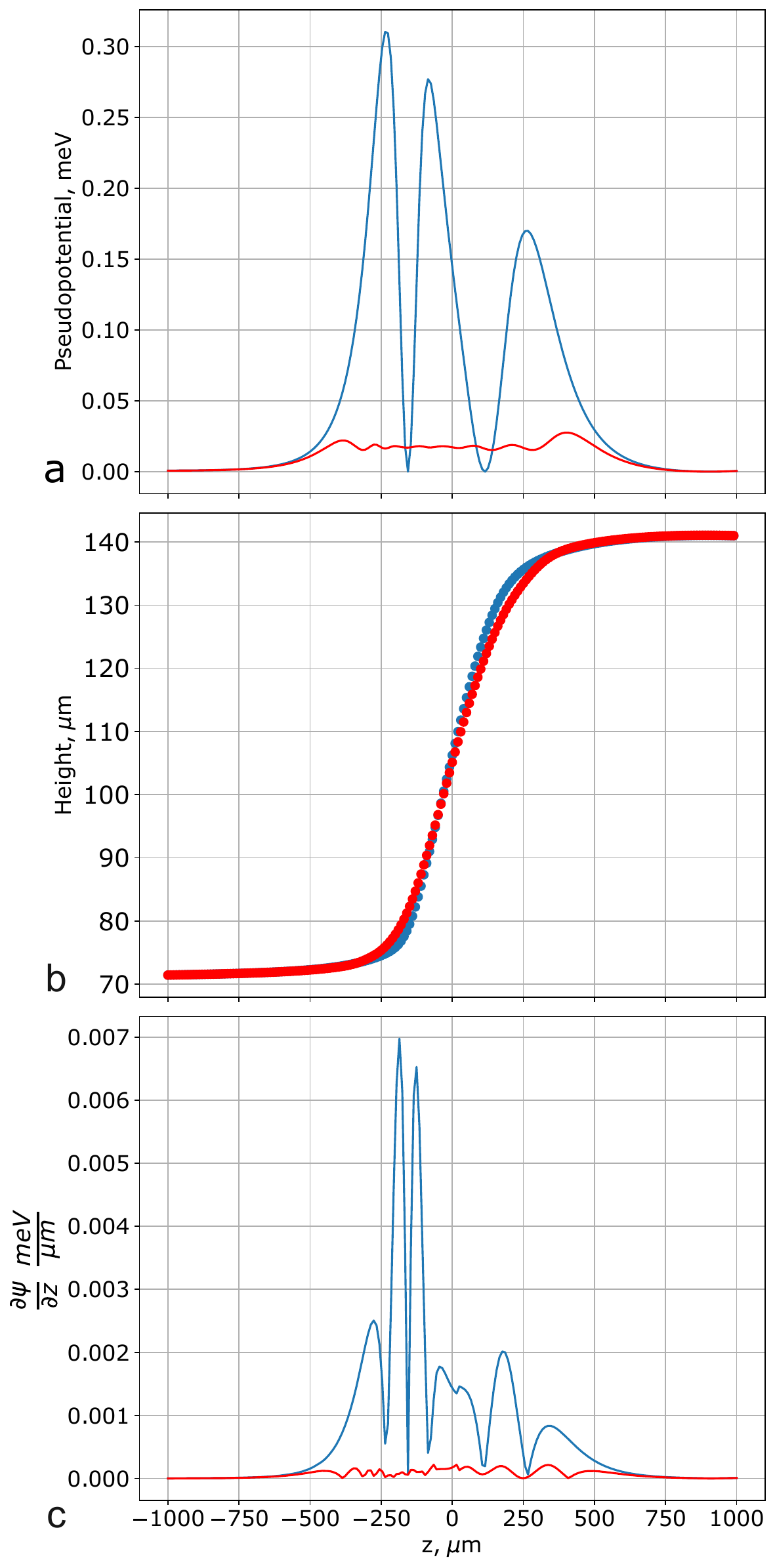}
    \caption{(a) Pseudopotential barrier along the ion transport path for optimized (red) and non-optimized (blue) transition regions; (b) Height of the pseudopotential minimum versus axial position; (c) Magnitude of the pseudopotential's first derivative.}
    \label{fig:suptus}
\end{figure}

The two-stage optimization process comprised:
\begin{enumerate}
  \item Individual objective optimization: each target function ($F_1-F_4$) was minimized separately to determine normalization coefficients, ensuring comparable scaling in the multi-objective function;
  \item Weighted-sum optimization: A composite objective function was minimized using empirically tuned weights $\sigma_i$.
\end{enumerate}
The optimization was performed using the Nelder--Mead algorithm. For the optimized design (Fig.~\ref{fig:dizain}), this yielded the following results: 10× reduction in pseudopotential magnitude (Fig.~\ref{fig:suptus}) and its gradient along the ion path for $<20\ \mu$m deviation of control points from initial x-positions.


\section{Ion elevator on a chip}
In contrast to the escalator, which achieves vertical transport through geometric variation of the electrode widths, the elevator approach shifts the RF null vertically by modifying the RF potential distribution. This requires no change in the electrode geometry along the axial direction, and thus introduces no pseudopotential barriers. The tradeoff is the need for an additional RF voltage source.

To enable a direct comparison of two elevator configurations, we analyze both methods using the same base trap geometry: a symmetric five-wire layout with equal electrode widths $a = b = c = 100\,\mu$m. Using the analytic 2D Laplace solution of House~\cite{house2008analytic}, we derive the trap parameters---confinement height, trap depth, and Mathieu $q$ parameter---as functions of the control voltage. All calculations are performed for $^{171}$Yb$^+$ with $V_\text{rf} = 100$\,V and $\Omega_\text{rf} = 2\pi \times 20$\,MHz.


\subsection{RF voltage applied to the central electrode}
The simplest elevator implementation applies an additional RF voltage $\alpha V_\text{rf}$ to the central ground electrode, where $\alpha$ is the voltage ratio (Fig.~\ref{fig:fig1}~(b)). For $\alpha > 0$, the central electrode is driven in phase with the RF rails; for $\alpha < 0$ (achieved by introducing a phase shift of $\pi$), the central electrode opposes them.

Setting the vertical component of the RF electric field to zero at the symmetry plane yields a closed-form expression for the confinement height:
\begin{equation}\label{eq:h_case1}
    h^2 = \frac{3a^2}{4}\cdot\frac{2 - 3\alpha}{2 + \alpha}\,,
\end{equation}
valid for $-2 < \alpha < 2/3$. At $\alpha = 0$ this reduces to the standard five-wire result $h_0 = a\sqrt{3}/2 \approx 86.6\,\mu$m~\cite{house2008analytic}. Positive $\alpha$ pulls the ion toward the surface, while negative $\alpha$ pushes it away.

\begin{figure}[htbp]
\centering
\includegraphics[width=0.47\textwidth]{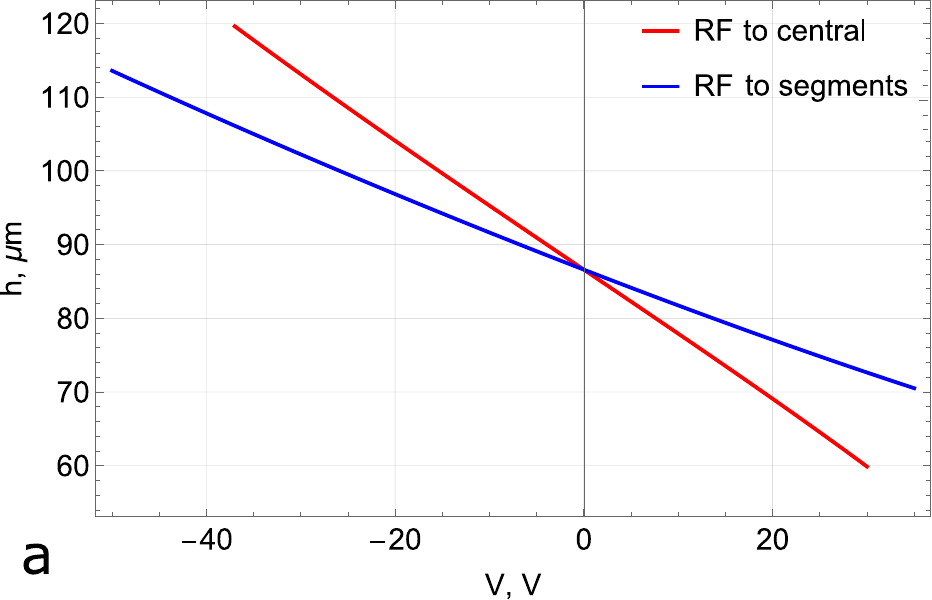}
\includegraphics[width=0.47\textwidth]{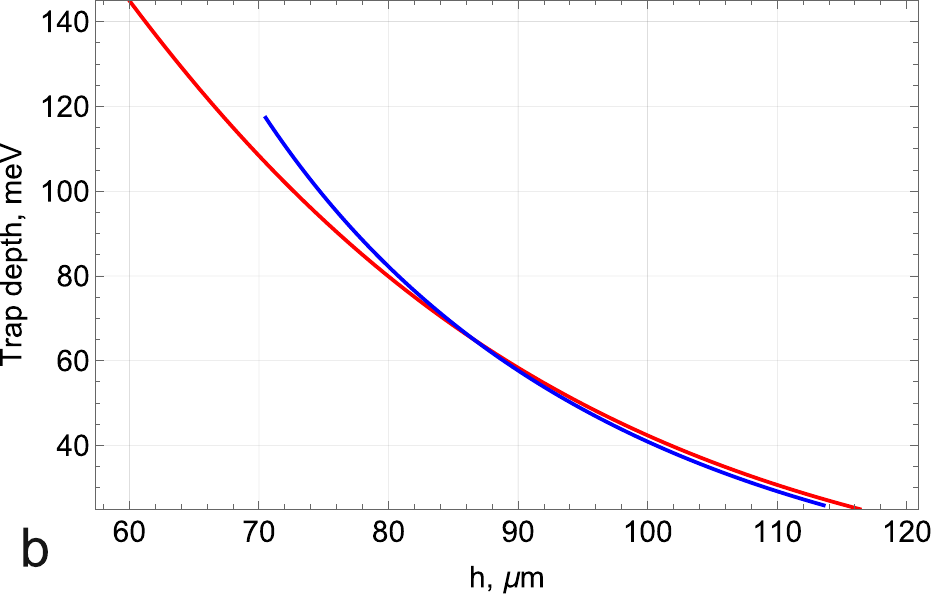}
\caption{Comparison of two elevator configurations for 
$a = b = c = 100\,\mu$m. (a)~Ion confinement height as a function of the control voltage. (b)~Trap depth as a function of confinement height. Only stable configurations with Mathieu parameter $q < 0.4$ and trap depth exceeding 25~meV are shown. Trapping may still be achievable beyond these bounds, but the pseudopotential approximation used here becomes unreliable and a full Mathieu stability analysis would be required.}
\label{fig:elev}
\end{figure}

The confinement height as a function of the control voltage is shown in Fig.~\ref{fig:elev}~(a). The dependence of the depth of the trap on the height is demonstrated in Fig.~\ref{fig:elev}~(b). Only parameter ranges that satisfy $q < 0.4$, and trap depths exceeding $25$~meV are shown. The height varies from approximately $60$ to $120\,\mu$m, allowing for nearly a twofold change in height with a single additional RF channel.

\subsection{RF voltage applied to segmented electrodes}

An alternative approach segments the central electrode into three parts (Fig.~\ref{fig:fig1}~(c)). The two outer segments, adjacent to the RF rails, are driven at voltage $\alpha V_\text{rf}$, while the central segment remains grounded. We consider the case in which all three parts have equal width.

Unlike the previous case, the three-term structure of the RF potential does not yield a simple closed-form expression for the null height. Nevertheless, the calculation remains analytic: all quantities are evaluated from the known potential~\cite{house2008analytic}. The results are shown alongside the first case in Fig.~\ref{fig:elev}.

The achievable height range is roughly half that of the first configuration 
for the same voltage swing, while the trap depth as a function 
of height is nearly identical for both configurations. However, the segmented design 
offers an additional degree of freedom: a DC voltage applied to the segments enables rotation 
of the principal axes of the trap \cite{niedermayr2015cryogenic} and micromotion 
compensation.



\section{Conclusion}
We have analyzed two complementary approaches to vertical ion positioning in surface Paul traps: the ``escalator'' and the ``elevator.'' Both are aimed at introducing controlled out-of-plane ion transport within the QCCD architecture.

The escalator connects two trapping regions with different inherent confinement heights through an optimized transition zone. Using a multi-objective optimization of the electrode boundary shape, we achieved a tenfold reduction in the pseudopotential barrier along the ion transport path, enabling adiabatic transport across a twofold height difference ($71$ to $141\,\mu$m) with minimal motional excitation. This method requires no additional voltage control and provides a passive, geometry-based division of the chip into zones of different confinement height.

The elevator adjusts the RF null height dynamically by applying 
a controllable RF voltage to additional electrodes. We analyzed 
two configurations on a common geometry ($a = b = c = 100\,\mu$m) 
using the analytic model: applying RF to the full central electrode, and driving segmented portions of the central electrode. The former provides a wider height tuning range for the same voltage swing. Both configurations yield nearly identical trap depth at a given confinement height and remain stable ($q < 0.4$) throughout the useful operating range.

The escalator and elevator methods address different operational needs and can be combined on a single chip. The escalator is suited for coarse height separation between functionally distinct zones---for instance, a low-height interaction zone optimized for quantum processing and a high-height memory zone with reduced surface noise. The elevator enables fine, continuous height adjustment within a given zone, as needed for precise alignment with an external cavity mode or for systematic studies of anomalous heating as a function of ion--surface distance. Together, these techniques extend the capabilities of the QCCD architecture into the third spatial dimension.


\bibliographystyle{quantum}
\bibliography{bib2}

@article{moses2023race,
  title={A race-track trapped-ion quantum processor},
  author={Moses, S.A. and Baldwin, C.H. and Allman, M.S. and Ancona, R. and Ascarrunz, L. and Barnes, C. and Bartolotta, J. and Bjork, B. and Blanchard, P. and Bohn, M. and others},
  journal={Phys. Rev. X},
  volume={13},
  number={4},
  pages={041052},
  year={2023},
  publisher={APS},
  doi={10.1103/PhysRevX.13.041052}
}

@article{blakestad2009high,
  title={High-fidelity transport of trapped-ion qubits through an X-junction trap array},
  author={Blakestad, R.B. and Ospelkaus, C. and VanDevender, A.P. and Amini, J.M. and Britton, J. and Leibfried, D. and Wineland, D.J.},
  journal={Phys. Rev. Lett.},
  volume={102},
  number={15},
  pages={153002},
  year={2009},
  publisher={APS},
  doi={10.1103/PhysRevLett.102.153002}
}

@article{DeCross2025computational,
  title={Computational Power of Random Quantum Circuits in Arbitrary Geometries},
  author={DeCross, M. and Haghshenas, R. and Liu, M. and Rinaldi, E. and Gray, J. and Alexeev, Y. and Baldwin, C.H. and Bartolotta, J.P. and Bohn, M. and Chertkov, E. and Cline, J. and Colina, J. and DelVento, D. and Dreiling, J.M. and Foltz, C. and Gaebler, J.P. and Gatterman, T.M. and Gilbreth, C.N. and Giles, J. and Gresh, D. and Hall, A. and Hankin, A. and Hansen, A. and Hewitt, N. and Hoffman, I. and Holliman, C. and Hutson, R.B. and Jacobs, T. and Johansen, J. and Lee, P.J. and Lehman, E. and Lucchetti, D. and Lykov, D. and Madjarov, I.S. and Mathewson, B. and Mayer, K. and Mills, M. and Niroula, P. and Pino, J.M. and Roman, C. and Schecter, M. and Siegfried, P.E. and Tiemann, B.G. and Volin, C. and Walker, J. and Shaydulin, R. and Pistoia, M. and Moses, S.A. and Hayes, D. and Neyenhuis, B. and Stutz, R.P. and Foss-Feig, M.},
  journal={Phys. Rev. X},
  volume={15},
  number={2},
  pages={021052},
  year={2025},
  publisher={APS},
  doi={10.1103/PhysRevX.15.021052}
}

@article{kielpinski2002architecture,
  title={Architecture for a large-scale ion-trap quantum computer},
  author={Kielpinski, D. and Monroe, C. and Wineland, D.J.},
  journal={Nature},
  volume={417},
  number={6890},
  pages={709--711},
  year={2002},
  publisher={Nature Publishing Group},
  doi={10.1038/nature00784}
}

@article{chiaverini2005surface,
  title={Surface-electrode architecture for ion-trap quantum information processing},
  author={Chiaverini, J. and Blakestad, R.B. and Britton, J. and Jost, J.D. and Langer, C. and Leibfried, D. and Ozeri, R. and Wineland, D.J.},
  journal={Quantum Inf. Comput.},
  volume={5},
  number={6},
  pages={419--439},
  year={2005},
  doi={10.48550/arXiv.quant-ph/0501147}
}

@article{Liu2014,
  title={A flexible optimization method for scaling surface-electrode ion traps},
  author={Liu, W. and Chen, S. and Wu, W.},
  journal={Appl. Phys. B},
  volume={117},
  number={4},
  pages={1149--1159},
  year={2014},
  publisher={Springer},
  doi={10.1007/s00340-014-5939-2}
}

@article{wright2013reliable,
  title={Reliable transport through a microfabricated X-junction surface-electrode ion trap},
  author={Wright, K. and Amini, J.M. and Faircloth, D.L. and Volin, C. and Doret, S.C. and Hayden, H. and Pai, C.S. and Landgren, D.W. and Denison, D. and Killian, T. and others},
  journal={New J. Phys.},
  volume={15},
  number={3},
  pages={033004},
  year={2013},
  publisher={IOP Publishing},
  doi={10.1088/1367-2630/15/3/033004}
}

@article{bowler2012coherent,
  title={Coherent diabatic ion transport and separation in a multizone trap array},
  author={Bowler, R. and Gaebler, J. and Lin, Y. and Tan, T.R. and Hanneke, D. and Jost, J.D. and Home, J.P. and Leibfried, D. and Wineland, D.J.},
  journal={Phys. Rev. Lett.},
  volume={109},
  number={8},
  pages={080502},
  year={2012},
  publisher={APS},
  doi={10.1103/PhysRevLett.109.080502}
}

@article{walther2012controlling,
  title={Controlling fast transport of cold trapped ions},
  author={Walther, A. and Ziesel, F. and Ruster, T. and Dawkins, S.T. and Ott, K. and Hettrich, M. and Singer, K. and Schmidt-Kaler, F. and Poschinger, U.},
  journal={Phys. Rev. Lett.},
  volume={109},
  number={8},
  pages={080501},
  year={2012},
  publisher={APS},
  doi={10.1103/PhysRevLett.109.080501}
}

@article{sterk2024multi,
  title={Multi-junction surface ion trap for quantum computing},
  author={Sterk, J.D. and Blain, M.G. and Delaney, M. and Haltli, R. and Heller, E. and Holterhoff, A.L. and Jennings, T. and Jimenez, N. and Kozhanov, A. and Meinelt, Z. and others},
  journal={arXiv preprint arXiv:2403.00208},
  year={2024},
  doi={10.48550/arXiv.2403.00208}
}

@article{kaushal2020shuttling,
  title={Shuttling-based trapped-ion quantum information processing},
  author={Kaushal, V. and Lekitsch, B. and Stahl, A. and Hilder, J. and Pijn, D. and Schmiegelow, C. and Bermudez, A. and M{\"u}ller, M. and Schmidt-Kaler, F. and Poschinger, U.},
  journal={AVS Quantum Sci.},
  volume={2},
  number={1},
  year={2020},
  publisher={AIP Publishing},
  doi={10.1116/1.5126186}
}

@article{kaufmann2017fast,
  title={Fast ion swapping for quantum-information processing},
  author={Kaufmann, H. and Ruster, T. and Schmiegelow, C.T. and Luda, M.A. and Kaushal, V. and Schulz, J. and von Lindenfels, D. and Schmidt-Kaler, F. and Poschinger, U.G.},
  journal={Phys. Rev. A},
  volume={95},
  number={5},
  pages={052319},
  year={2017},
  publisher={APS},
  doi={10.1103/PhysRevA.95.052319}
}

@article{ramette2022any,
  title={Any-to-any connected cavity-mediated architecture for quantum computing with trapped ions or Rydberg arrays},
  author={Ramette, J. and Sinclair, J. and Vendeiro, Z. and Rudelis, A. and Cetina, M. and Vuleti{\'c}, V.},
  journal={PRX Quantum},
  volume={3},
  number={1},
  pages={010344},
  year={2022},
  publisher={APS},
  doi={10.1103/PRXQuantum.3.010344}
}

@article{takahashi2020strong,
  title={Strong coupling of a single ion to an optical cavity},
  author={Takahashi, H. and Kassa, E. and Christoforou, C. and Keller, M.},
  journal={Phys. Rev. Lett.},
  volume={124},
  number={1},
  pages={013602},
  year={2020},
  publisher={APS},
  doi={10.1103/PhysRevLett.124.013602}
}

@article{kassa2018precise,
  title={Precise positioning of an ion in an integrated Paul trap-cavity system using radiofrequency signals},
  author={Kassa, E. and Takahashi, H. and Christoforou, C. and Keller, M.},
  journal={J. Mod. Opt.},
  volume={65},
  number={5--6},
  pages={520--528},
  year={2018},
  publisher={Taylor \& Francis},
  doi={10.1080/09500340.2017.1406158}
}

@article{brownnutt2015ion,
  title={Ion-trap measurements of electric-field noise near surfaces},
  author={Brownnutt, M. and Kumph, M. and Rabl, P. and Blatt, R.},
  journal={Rev. Mod. Phys.},
  volume={87},
  number={4},
  pages={1419--1482},
  year={2015},
  publisher={APS},
  doi={10.1103/RevModPhys.87.1419}
}

@article{house2008analytic,
  title={Analytic model for electrostatic fields in surface-electrode ion traps},
  author={House, M.G.},
  journal={Phys. Rev. A},
  volume={78},
  number={3},
  pages={033402},
  year={2008},
  publisher={APS},
  doi={10.1103/PhysRevA.78.033402}
}

@article{leibfried2003quantum,
  title={Quantum dynamics of single trapped ions},
  author={Leibfried, D. and Blatt, R. and Monroe, C. and Wineland, D.},
  journal={Rev. Mod. Phys.},
  volume={75},
  number={1},
  pages={281},
  year={2003},
  publisher={APS},
  doi={10.1103/RevModPhys.75.281}
}

@article{kim2010surface,
  title={Surface-electrode point Paul trap},
  author={Kim, T.H. and Herskind, P.F. and Kim, T. and Kim, J. and Chuang, I.L.},
  journal={Phys. Rev. A},
  volume={82},
  number={4},
  pages={043412},
  year={2010},
  publisher={APS},
  doi={10.1103/PhysRevA.82.043412}
}

@article{van2016integrated,
  title={An integrated mirror and surface ion trap with a tunable trap location},
  author={Van Rynbach, A. and Maunz, P. and Kim, J.},
  journal={Appl. Phys. Lett.},
  volume={109},
  number={22},
  year={2016},
  publisher={AIP Publishing},
  doi={10.1063/1.4970542}
}

@article{boldin2018measuring,
  title={Measuring anomalous heating in a planar ion trap with variable ion-surface separation},
  author={Boldin, I.A. and Kraft, A. and Wunderlich, C.},
  journal={Phys. Rev. Lett.},
  volume={120},
  number={2},
  pages={023201},
  year={2018},
  publisher={APS},
  doi={10.1103/PhysRevLett.120.023201}
}

@article{an2018surface,
  title={Surface trap with dc-tunable ion-electrode distance},
  author={An, D. and Matthiesen, C. and Abdelrahman, A. and Berlin-Udi, M. and Gorman, D. and M{\"o}ller, S. and Urban, E. and H{\"a}ffner, H.},
  journal={Rev. Sci. Instrum.},
  volume={89},
  number={9},
  year={2018},
  publisher={AIP Publishing},
  doi={10.1063/1.5046527}
}

@article{hughes2025trapped,
  title={Trapped-ion two-qubit gates with> 99.99\% fidelity without ground-state cooling},
  author={Hughes, A.C. and Srinivas, R. and L{\"o}schnauer, C.M. and Knaack, H.M. and Matt, R. and Ballance, C.J. and Malinowski, M. and Harty, T.P. and Sutherland, R.T.},
  journal={arXiv preprint arXiv:2510.17286},
  year={2025},
  doi={10.48550/arXiv.2510.17286}
}

@article{kolachevsky2025quantum,
  title={Quantum computing with trapped ions: principles, achievements, and prospects},
  author={Zalivako, I.V. and Semenin, N.V. and Zhadnov, N.O. and Galstyan, K.P. and Kamenskikh, P.A. and Smirnov, V.N. and Korolkov, A.E. and Sidorov, P.L. and Borisenko, A.S. and Anosov, Yu.P. and Semerikov, I.A. and Khabarova, K.Yu. and Kolachevsky, N.N.},
  journal={Phys.-Usp.},
  volume={68},
  number={6},
  pages={552--583},
  year={2025},
  publisher={Russian Academy of Sciences},
  doi={10.3367/UFNe.2024.12.039884}
}

@article{zhang2022optimization,
  title={Optimization and implementation of a surface-electrode ion trap junction},
  author={Zhang, C. and Mehta, K.K. and Home, J.P.},
  journal={New J. Phys.},
  volume={24},
  number={7},
  pages={073030},
  year={2022},
  publisher={IOP Publishing},
  doi={10.1088/1367-2630/ac7db6}
}

@misc{ransford2025helios,
  title={Helios: A 98-qubit trapped-ion quantum computer},
  author={Quantinuum},
  year={2025},
  eprint={2511.05465},
  archivePrefix={arXiv},
  primaryClass={quant-ph},
  url={https://arxiv.org/abs/2511.05465}
}

@article{channa2019surface,
  title={Surface Ion Trap Designs for Vertical Ion Shuttling},
  author={Channa, M.Y. and Nizamani, A.H. and Saleem, H. and Bhutto, W.A. and Soomro, A.M. and Soomro, M.Y.},
  journal={IJCSNS},
  volume={19},
  number={4},
  pages={264},
  year={2019},
  url={http://paper.ijcsns.org/07_book/201904/20190436.pdf}
}

@phdthesis{niedermayr2015cryogenic,
  title={Cryogenic surface ion traps},
  author={Niedermayr, M.},
  year={2015},
  school={University of Innsbruck (Austria)}
}

@article{gerasin2024optimized,
  title={Optimized surface ion trap design for tight confinement and separation of ion chains},
  author={Gerasin, I. and Zhadnov, N. and Kudeyarov, K. and Khabarova, K. and Kolachevsky, N. and Semerikov, I.},
  journal={Quantum Reports},
  volume={6},
  number={3},
  pages={442--451},
  year={2024},
  publisher={MDPI},
  doi={10.3390/quantum6030029}
}

\onecolumn
\appendix
\section{Initial trap design}

\begin{figure}[h]
    \centering
    \begin{minipage}{0.48\textwidth}
        \includegraphics[width=\textwidth]{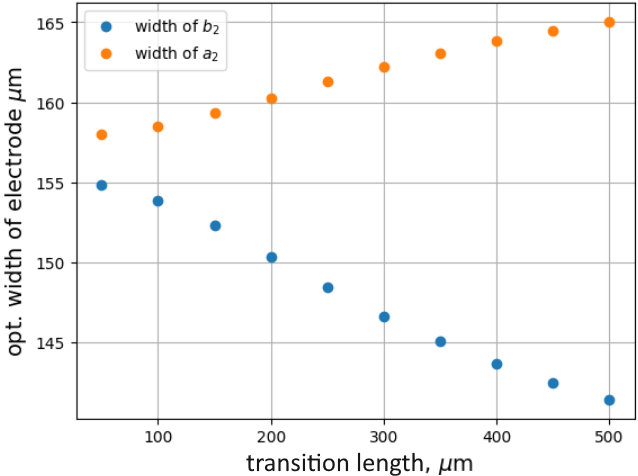}
        \caption*{a}
    \end{minipage}
    \begin{minipage}{0.49\textwidth}
        \includegraphics[width=\textwidth]{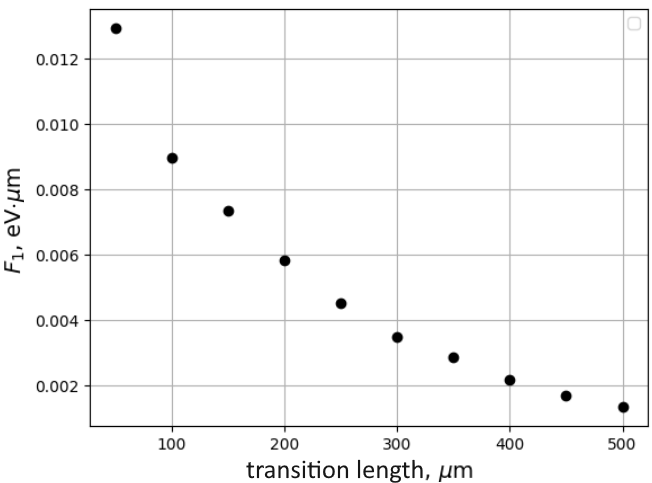}
        \caption*{b}
    \end{minipage}
    \caption{(a)~Optimal widths of the central GND electrode $a_2$ and RF electrodes $b_2$ of the second trap as functions of the transition region length. (b)~Corresponding values of the objective function $F_1$.}
    \label{fig:2}
\end{figure}

Before optimizing the transition region shape, we first determine the electrode geometry of the second (taller) trap and the transition length. For each candidate transition length, the widths of the GND and RF electrodes of the second trap were optimized using the Nelder--Mead algorithm with the objective function
\begin{equation}
    F_1 = \int_{-l}^{l}\psi(y|_{\psi_{\min}})\,dz\,,
\end{equation}
which minimizes the integrated pseudopotential along the ion transport path. The resulting optimal electrode widths and the corresponding values of $F_1$ are shown in Fig.~\ref{fig:2} as functions of the transition length.

Increasing the transition length reduces both the peak and the mean pseudopotential barrier along the ion path. However, an excessively long transition is impractical, as it increases the transport distance and time, leading to additional motional heating. As a compromise, we selected a transition length of $D = 300\,\mu$m for subsequent connector optimization described in Section~2.2.

\end{document}